\documentclass[pra,aps,amssymb,twocolumn,noshowpacs,hyperref,superscriptaddress]{revtex4-1}
\usepackage{color}
\usepackage{graphicx}
\usepackage{hyperref}
\usepackage{amsmath}
\usepackage{latexsym}
\usepackage{amssymb}
\usepackage{mathrsfs}
\usepackage{mathtools}
\usepackage{layout}
\usepackage{verbatim}
\usepackage{gensymb}
\usepackage{bm}
\usepackage{amsfonts,epsfig}

\begin{document}
\title{The fractional Chern insulator with Rydberg-dressed neutral atoms}
\date{\today}
\author{Yang Zhao}
\affiliation{ Hebei Normal University, Shijiazhuang, Hebei 050024, China}
\affiliation{ Northwestern Polytechnical University, Xi'an, 710072, China}
\author{Xiao-Feng Shi}
\affiliation{School of Physics, Xidian University, Xi'an 710071, China}


\begin{abstract}
Topological nontrivial bands can be realized via Rydberg-dressed neutral atoms. We propose a two-dimensional hard-core boson model with a topological ground enrgy flat band on a honeycomb lattice, where the particle hopping is realized via van der Waals interaction that exchanges the Rydberg states of two interacting atoms, while nonzero phases associated with hopping is created by transferring the optical phase of laser fields to the atomic pair wave function. Using exactly diagonalization and infinite density matrix renormalization group simulation, we find in the system a fractional Chern insulator phase with a Chern number $C=1/2$, which can persist in the presence of weak many-body interactions. Our studies indicate that fractional Chern insulators can be studied with neutral-atom arrays.
\end{abstract}

\maketitle

\section{introduction}
Quantum simulation of condensed-matter physics via atomic, molecular, and optical methods has become an interesting topic recently~\cite{Feynman1982,Bloch2008,Bloch2012,Georgescu2014}. Among various phases of condensed matter, topological phases~\cite{Klitzing1980,Haldane1988} are especially interesting not only because of their relevance for fundamental condensed matter physics~\cite{Qi2009,Hasan2010,Qi2011}, but also due to their potential in quantum computing~\cite{Kitaev2006,Nayak2008}. Natural topological phases of matter are not easily manipulated, but it is possible to engineer spin-orbit coupling~\cite{Goldman2010,Liu2014} or pseudo-magnetic field~\cite{Sorensen2005,Wu2008,Stanescu2009,Tang2011,Sun2011,Wang2011,Neupert2011,Yao2012,Yao2013} to prepare topological phases, especially with neutral atoms~\cite{Samajdar2021,Browaeys2020,Adams2020,Wu2021,Morgado2021,PhysRevLett.124.140401,PhysRevA.101.043618,PhysRevX.10.021041,PhysRevX.11.031005,PhysRevLett.128.017601,PhysRevLett.127.263004}.

In this work, we show that it is possible to use van der Waals interaction between Rydberg atoms~\cite{Gallagher1994,Walker2008,Saffman2010,Wu2022} in a two-dimensional lattice to simulate a topological Chern insulator~\cite{Thouless1982,Khazali2022} with a fractional Chern number $C=1/2$. The method depends on an effective magnetic field created by off-resonantly addressing of Rydberg states as proposed in Ref.~\cite{Shi2018}. A Rydberg state refers to a state where the atomic energy is much higher than its ground-state energy. Because the electron of a Rydberg atom is extended far from its nucleus, the dipole-dipole interaction between two Rydberg atoms is much larger than that between two corresponding ground-state atoms~\cite{Shi2021}. Due to the strong dipole-dipole interaction, there has been intense efforts directed at the search for exotic effective magnetic phases of atoms in optical lattices via Rydberg interactions~\cite{Qian2012,Honing2013,Hoening2014,Glaetzle2015,VanBijnen2015,Kiffner2016,Weber2018}. The possibility to realize topological states in a one-dimensional system by Rydberg atoms was demonstrated in~\cite{DeLeseleuc2019}.

The two-dimensional topological state in our method results from an off-resonant addressing of ground states to Rydberg states~\cite{Santos2000,Bouchoule2002,Pupillo2010,Henkel2010,Mattioli2013,Macri2014,Glaetzle2014,Glaetzle2015,VanBijnen2015,Potirniche2017}, so that the effective ground-state atoms inherent interactions from the Rydberg states. Our goal is to design a tight-binding Hamiltonian which has one or more bands that are topological nontrivial, i.e., possessing nonzero Chern numbers when the time-reversal symmetry is absent. The first step to achieve this is to build up quasi-particle hopping between atoms, and the second is, as one choice, to induce a pseudo-magnetic field so that the time-reversal symmetry is broken. In our model, The hopping between two lattice sites is realized by the exchange interaction~\cite{Afrousheh2004,Gunter2013,DeLeseleuc2017,Shi2021} when two interacting atoms are in different Rydberg levels, while the pseudo-magnetic field can be created via the phase difference of lasers upon the two atoms~\cite{Shi2018}. Since the lattice constant of our model is on the order of $10\mu$m, the benefit of realizing topological band structure via van der Waals interactions is that accurate control of a single atom in the lattice can be easily carried out~\cite{Isenhower2010}, thus rendering easier ways to, for instance, study impurities~\cite{Liu2009} in topological phases and manipulate excitations for topological quantum computation~\cite{You2010,Laflamme2014,PhysRevX.11.031005}. Compared to other methods by using direct dipolar exchange interactions of Rydberg atoms~\cite{Kiffner2016,Weber2018}, the Rydberg dressing in this work renders long coherence times for the quasiparticles~\cite{Santos2000,Bouchoule2002,Pupillo2010,Henkel2010,Mattioli2013,Macri2014,Glaetzle2014,Glaetzle2015,VanBijnen2015,DeLeseleuc2019}.

The remainder of this article is outlined as follows. In Sec.~\ref{sec02}, we address the technical details of the off-resonant addressing of ground states to Rydberg states and deduce the effective Hamiltonian of the system which describes interacting hard-core bosons on a honeycomb lattice. In Sec.~\ref{sec03}, we prove the existence of a fractonal Chern insulator (FCI) phase and propose the parameter settings for the realiaztion of the FCI phase in this model by using the exact diagonalization (ED) and the infinite-size variant of the Density Matrix Renormalization Group (iDMRG) method, and discuss the phase transitions between the FCI phase and other topological trivial phases. Section~\ref{sec04} gives a brief summary.

\section{A honeycomb Rydberg-dressed systems with effective magnetic fields}\label{sec02}
\begin{figure}
\includegraphics[width=1.0\linewidth]
{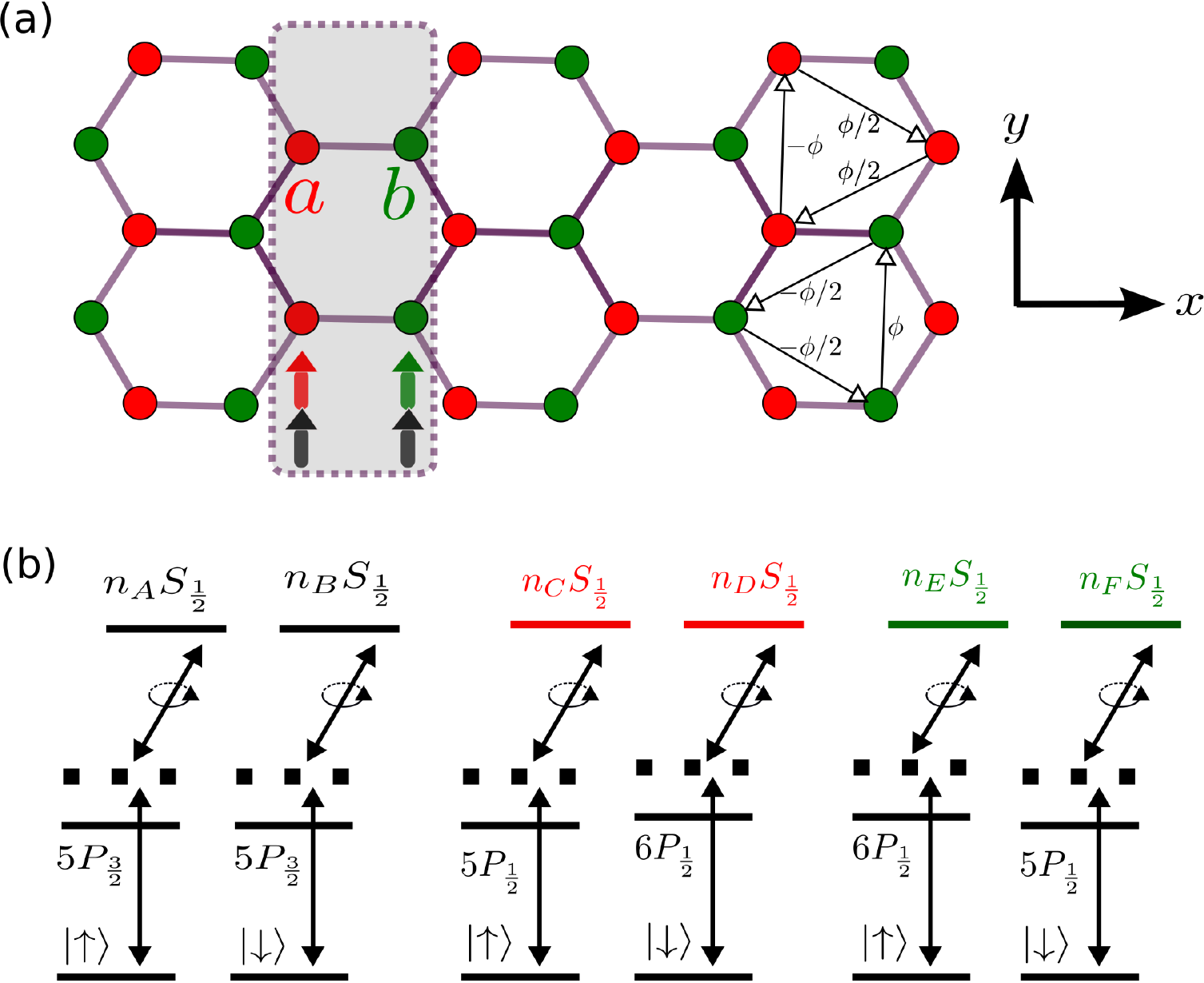}
 \caption{(a) Quantum emulation of a two-dimensional hard-core boson-Hubbard model in a honeycomb lattice. The numbers $\pm\phi$ and $\pm\phi/2$ show the phase changes when hard-core bosons hop from one site to another, shown by the arrows. (b) Optical admixing of Rydberg states with two ground states for each atom in the optical lattice. For each atom, two chosen ground states are excited by two-photon transitions. Each two-photon process is realized by a $\mathbf{z}$-polarized laser~(lower transition) that travels along $\mathbf{y}$ and a right-hand polarized laser~(upper transition) that travels along $\mathbf{z}$. \label{fig1}}
\end{figure}

We focus on a honeycomb optical lattice, where the two sublattices are labeled as a and b, as shown in Fig.~\ref{fig1}(a). Trapped at each site of the lattice is one $^{87}Rb$ atom that is optically excited off-resonantly. Each atom is prepared in a superposition state of two hyperfine ground states $|\uparrow  \rangle$ and $|\downarrow  \rangle$. These two states can be chosen as $|$$5S_{1/2},F=1(2),m_F=1\rangle $, where $F$ and $m_F$ denote hyperfine and magnetic quantum numbers, respectively. The nearest neighboring couplings are realized by coupling the two ground states to two Rydberg levels $(n_AS_{1/2},n_BS_{1/2}) $ for all sites, where $n_B-n_A=1$. All the coupled Rydberg states have electron and nuclear spin states $1/2$ and $3/2$, respectively, thus these quantum numbers are suppressed when writing the Rydberg atomic states. The van der Waals interaction can drive the two-atom state $|n_AS_{1/2}n_BS_{1/2}\rangle$ to $|n_BS_{1/2}n_AS_{1/2}\rangle$~\cite{Shi2021}, where the symbols inside the ket on the left and right denote states of the two neighboring atoms. When addressed to the ground states, this type of exchange process will give the exchange interaction $|\uparrow\downarrow\rangle\leftrightarrow|\downarrow\uparrow\rangle$ between two neighboring atoms. Since the van der Waals interaction scales as $1/r^6$, with $r$ the distance between two atoms, the addressing of the ground states to these two levels will only lead to nearest-neighboring effective interaction, while next-nearest neighboring interaction can be ignored. In order to switch on next-nearest neighboring couplings, the ground states are also coupled to two other pairs of Rydberg levels $(n_CS_{1/2},n_DS_{1/2}) $ and $(n_ES_{1/2},n_FS_{1/2})$ for the two sublattices, respectively, where $n_D-n_C=n_F-n_E=1$. Since the distance between two next-nearest neighboring sites are $\sqrt3$ larger compared to that between two nearest neighbors, $n_C,\cdots, n_F$ shall be larger than $n_A$ and $n_B$, so that the coupling between two nearest sites and that between two next-nearest sites are comparable in magnitude.

The comparable couplings for both nearest and next-nearest neighbors are realized by choosing appropriate Rydberg levels. Even the blockade interaction occurs between two atoms in any Rydberg levels, the exchange interaction occurs only for a pair of levels $n_\alpha, n_\beta$ when $|n_\alpha-n_\beta|$ is small. As shown in Ref.~\cite{SBK2014}, the van der Waals interaction that exchanges the two principal quantum numbers $n_\alpha$ and $ n_\beta$ is negligible when $|n_\alpha-n_\beta|>2$. When we choose the condition $n_C-n_B, n_E-n_D >2$, the exchange process between $(n_AS_{1/2},n_BS_{1/2}) $, $(n_CS_{1/2},n_DS_{1/2}) $, or $(n_ES_{1/2},n_FS_{1/2})$ is the dominant exchange interaction. This design can give us a large degree of freedom to adjust parameters for simulating different possible magnetic phases~\cite{Glaetzle2015,VanBijnen2015}. Beside of the interaction that exchanges the principal quantum numbers of two atoms, there is a residual interaction that changes the total electron spin of two atoms, but its magnitude is several orders smaller than that of the former process, thus can be ignored.

The time-reversal symmetry is broken by introducing a nonzero phase to the next-nearest neighboring hopping through the laser excitation. To understand such a phase term for the hopping, we can look at the various optical excitations of the neutral atoms. For all the two-photon transitions depicted in Fig~\ref{fig1}(b), the lower transitions are through linearly polarized lasers, whose electric vectors are polarized along $\mathbf{z}$, while the upper transitions happen through right-hand polarized laser fields. Now the two-dimensional lattice lies in the $x-y$ plane, then it is convenient to choose $z=0$ for the plane where the optical lattice lies. As a result, the phase term with a chosen two-photon Rabi frequency is solely determined by the lower Rabi transition. For sublattice a, we denote the phase term accompanying the transition from $|\uparrow(\downarrow)\rangle$ to $|n_{C(D)}S_{1/2}\rangle$ as $\phi(\mathbf{r}_i) [\varphi(\mathbf{r}_i)]$ by the two photon laser excitation as in Fig~\ref{fig1}(b). By performing perturbation calculation as from Refs.~\cite{Shavitt1980,Roland2003}, the phase term appeared in $|\uparrow\downarrow\rangle\langle\downarrow\uparrow|_{ij}$ for the effective Hamiltonian reads $\phi_{ij}=\phi(\mathbf{r}_i)-\phi(\mathbf{r}_j) - \varphi(\mathbf{r}_i)+\varphi(\mathbf{r}_j)$. In the setup of Fig~\ref{fig1}(b), the two-photon transitions to $(n_CS_{1/2},n_DS_{1/2}) $ are via $5P_{1/2}$ and $6P_{1/2}$ intermediate levels, giving $\phi_{ij}\sim (E_{5P}-E_{6P})/(\hbar c) \mathbf{\hat y}\cdot (\mathbf{r}_i-\mathbf{r}_j)$, with $\hbar$ the reduced Planck constant, $c$ the speed of light in vacuum, $E_{5(6)P}$ the energy of the atomic level. Because the $5P_{1/2}$ and $6P_{1/2}$ levels have an energy difference of about $333.9\times 2\pi$THz~\cite{Sansonetti2006}, a significant phase $\phi_{ij}$ can appear. More details about such phases can be found in Ref.~\cite{Shi2018}.

The effective Hamiltonian is derived in a perturbative method. When the two-photon detuning is large compared to the two-photon Rabi frequency, the coupling between the ground and the excited states are removed effectively by a canonically transformation~\cite{Shavitt1980}. Details of the derivation can be found in Ref.~\cite{Shi2018}. Up to fourth order, the effective Hamiltonian is given by
\begin{eqnarray}
\hat{H}_{\text{eff}}^{(4)}  &=& \sum_{\alpha,\beta,\gamma,\epsilon=\uparrow,\downarrow}\mathsf{H}_{\alpha\beta,\gamma\epsilon} |\alpha\beta\rangle\langle \gamma\epsilon|, \label{eq02}
\end{eqnarray}
where $\mathsf{H}_{\alpha\beta,\gamma\epsilon}$ is a function of the Rabi frequency, detuning, and Rydberg interaction~\cite{Shi2018}.
The total effective spin of the system $\sum_i |\uparrow\rangle\langle\uparrow|$ is conserved, which motivates us to define hard-core bosons. We note that for a system with dipolar interactions, such a conservation of the total effective spin renders a system of hard-core boson Hubbard model that can be topologically nontrivial~\cite{Yao2012,Yao2013,Peter2015,Maghrebi2015}. Defining $b_i^\dag\equiv (|\uparrow\rangle\langle\downarrow|)_i$ and $\hat  n_i =b_i^\dag b_i $ for each site $i$, the system has an effective Hamiltonian
 \begin{eqnarray}
\hat{H}_{\text{b}}
&= &  \sum_{\langle i,j\rangle}  \left(t_{ij} \hat b_i^\dag \hat b_j +\text{h.c.}\right) +  \sum_{\langle i,j\rangle}U_{ij} \hat  n_i \hat n_j +\sum_i \mu_i  \hat n_i,\nonumber\\
 \label{Hubbard1}
\end{eqnarray}
where $\mu_{i} = \sum_{j\neq i} [( \mathsf{H}_{\uparrow\downarrow, \uparrow\downarrow}+ \mathsf{H}_{\downarrow\uparrow, \downarrow\uparrow})/2 - \mathsf{H}_{\downarrow\downarrow, \downarrow\downarrow } ]_{i,j},U_{ij}=[\mathsf{H}_{ \uparrow\uparrow , \uparrow\uparrow  } + \mathsf{H}_{\downarrow\downarrow, \downarrow\downarrow }- \mathsf{H}_{\uparrow\downarrow, \uparrow\downarrow} - \mathsf{H}_{\downarrow\uparrow, \downarrow\uparrow} ]_{i,j} $, and $t_{ij} =\mathsf{H}_{\uparrow\downarrow, \downarrow\uparrow} $. For convenience, we label the nearest-neighboring~(N) and next-nearest-neighboring~(NN) hoppings between the two sublattice as $t_{\text{\tiny{N}}}^{(ab)}$ and $t_{\text{\tiny{NN}}}^{(ab)}$, respectively. The magnitudes of the NN hoppings in sublattice a and b are $t_{\text{\tiny{NN}}}^{(aa)}$ and $t_{\text{\tiny{NN}}}^{(bb)}$, respectively, while the phases accompanying these hoppings are illustrated in Fig.~\ref{fig1}. The chemical potential and the NN many-body interaction are $\mu^{(x)}$ and $U_{\text{\tiny{NN}}}^{(x)}$ for sublattice x, where $x=$a or b, while the nearest-neighboring many-body interaction is $U_{\text{\tiny{N}}}^{(ab)}$. These interactions are nonzero in general, indicating that it is possible to realize interesting many-body phases through van der Waals interactions of neutral atoms.

Equation~(\ref{Hubbard1}) is analogous to the Haldane-Bose-Hubbard model\cite{Haldane1988}\cite{Wang2011} filled with hard-core bosons except for the different phase factors pertaining to the next-nearest hopping $t_{\text{\tiny{NN}}}^{(ab)}$ that generate zero net flux threading within one unit cell, as shown in Fig.~\ref{fig1}(a). We will name our model described by Eq.~(\ref{Hubbard1}) as the untypical Haldane-Bose-Hubbard (UHB) model in the following context for convenience in comparison with the usual Haldane-Bose-Hubbard model.
Previous studies\cite{Wang2011}\cite{Luo2020} of the Haldane-Bose-Hubbard model have shown that a fractional Chern insulator (FCI) phase with a fractional Chern number $C=1/2$ can be realized in such a model with parameter settings under which a flat ground energy band is achieved. Such a model is termed as topological flat band model (TFB). Consequently, it is natural to expect that such a FCI phase may also be realized in the UHB model.

\begin{figure}
\includegraphics[width=0.8\linewidth]
{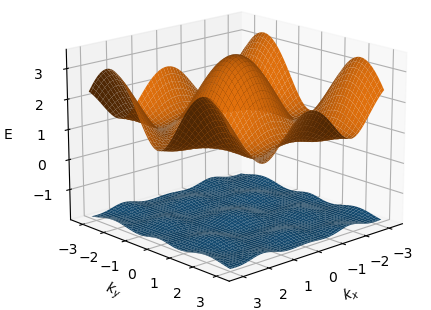}
 \caption{The band structure of Eq.~(\ref{Hubbard1}) for
which the flatness ratio for the ground energy band is 0.099, and $t_{\text{\tiny{N}}}=$-1, $\phi=$8.37700016, $t_{\text{\tiny{NN}}}=$-0.31, $t_{\text{\tiny{NN}}}=$0.14.\label{fig2}}
\end{figure}

Since the existence of flat bands is the prerequisite for the FCI phase to appear, it is useful to show a set of parameters for Eq.~(\ref{Hubbard1}) to support TFB. The flat ground energy band is described by the flatness ratio which is defined as $\delta/\Delta$\cite{Neupert2011}, where $\delta$ is the band width of the ground energy band and $\Delta$ denotes the energy gap above the ground energy band. In the non-interacting case, Eq.~(\ref{Hubbard1}) can be directly diagonalized by means of Fourier transformation which gives $H_{\text{k}} = d_0(\mathbf{k}) +  \sum_{j=1}^3 d_j(\mathbf{k}) \sigma^j$, where $\sigma^j$ is the $j$th Pauli matrix, and
\begin{eqnarray}
d_0&=& 2t_2(\cos\phi\cos\mathbf{k}\mathbf{v}_1+\cos\frac{\phi}{2}\cos\mathbf{k}\mathbf{v}_2+\cos\frac{\phi}{2}\cos\mathbf{k}\mathbf{v}_3)\nonumber\\
d_1&=&t(\cos\mathbf{k}\mathbf{s}_1+\cos\mathbf{k}\mathbf{s}_2+\cos\mathbf{k}\mathbf{s}_3)\nonumber\\
d_2&=&t(\sin\mathbf{k}\mathbf{s}_1+\sin\mathbf{k}\mathbf{s}_2+\sin\mathbf{k}\mathbf{s}_3)\nonumber\\
d_3&=&2t_2(-\sin\phi\sin\mathbf{k}\mathbf{v}_1+\sin\frac{\phi}{2}\sin\mathbf{k}\mathbf{v}_2+\sin\frac{\phi}{2}\sin\mathbf{k}\mathbf{v}_3).\nonumber
\end{eqnarray}
Here $\mathbf{v}_1=(\sqrt{3},0)$, $\mathbf{v}_{2,3}=(-\sqrt{3}/2,\pm3/2)$ and $\mathbf{s}_{1,2}=(\pm\sqrt{3}/2,1/2)$, $\mathbf{s}_{3}=(0,-1)$. Then, the two eigenvalues of $H_{\text{k}}$ can be written as $\varepsilon_{\pm,\mathbf{k}}=d_0(\mathbf{k})\pm\sqrt{d^2_1(\mathbf{k})+d^2_2(\mathbf{k})+d^2_3(\mathbf{k})}$. We can read out the band width of the lower band $\varepsilon_{-,\mathbf{k}}$: $\delta=\text{max}[\varepsilon_{-,\mathbf{k}}]-\text{min}[\varepsilon_{-,\mathbf{k}}]$, and the energy gap: $\Delta=\text{min}[\varepsilon_{+,\mathbf{k}}]-\text{max}[\varepsilon_{-,\mathbf{k}}]$. The band flatness ratio $\delta/\Delta$ can be obtained numerically and the optimal value for it is fixed at about 0.099 with the parameter setting: $t_{\text{\tiny{N}}}^{(ab)}=$-1, $\phi=$8.37700016, $t_{\text{\tiny{NN}}}^{(ab)}=$-0.31, $t_{\text{\tiny{NN}}}^{(aa) or (bb)}=$0.14, as shown in Fig~\ref{fig2}. Here, by following Ref\cite{Wang2011}, we have taken the next-next-nearest neighbour hopping terms into consideration in order to obtain a more flatter band structure. During the deduction, we have removed all the superscripts of the hopping amplitudes $t$s and coupling strength $U$s for conciseness and will follow this convention in the following context.

Next, in the following section, we will search for the expected FCI phase and its rubustness to the finite coupling $U_{\text{\tiny{N}}}^{(ab)}$ and $U_{\text{\tiny{NN}}}^{(a) or (b)}$ in the above TFB UHB model and briefly discuss the non-topological phases exhibited in its phase diagram.
\section{Numerical results}\label{sec03}
In this section, we perform the ED and iDMRG calculations to ascertain the possible phases exhibited by the UHB model with a flat ground energy band and demonstrate the existence of a FCI phase in the phase diagram of the model with respect to $U_{\text{\tiny{NN}}}$ and $U_{\text{\tiny{N}}}$. The stability of this FCI phase against the presence of finite interactions and the phase transitions driven by the ratio of $U_{\text{\tiny{NN}}}/U_{\text{\tiny{N}}}$ are also studied numerically.

To facilitate the numerical calculation and utilize the translational symmetry of the UHB model, we roll the honeycomb lattice shown in Fig.~\ref{fig1}(a) as an infinite cylinder along the $x$ axis while with a finite circumstance $L_y$ unit cells along the $y$ direction. For the ED calculation, we will consider a 3$\times$4 ($L_y\times L_x$) lattice with periodical boundary along both of the space directions.

We concentrate on the half-filling Hilbert subspace with the filling factor $\nu=N_b/N_{cell}$=1/2, here $N_b$ is the particle number and $N_{cell}$ is the total number of unit cells of the lattice. It has been verified for the TFB Haldane-Bose-Hubbard model that the ground state with $\nu=1/2$ in the weak coupling case is a $1/2$ bosonic fractional Chern insulator\cite{Wang2011}. Therefore, a $1/2$ bosonic FCI state is also naturally expected in our model due to its analogy to the usual Haldane-Bose-Hubbard model.
\begin{figure}
\includegraphics[width=1.0\linewidth]
{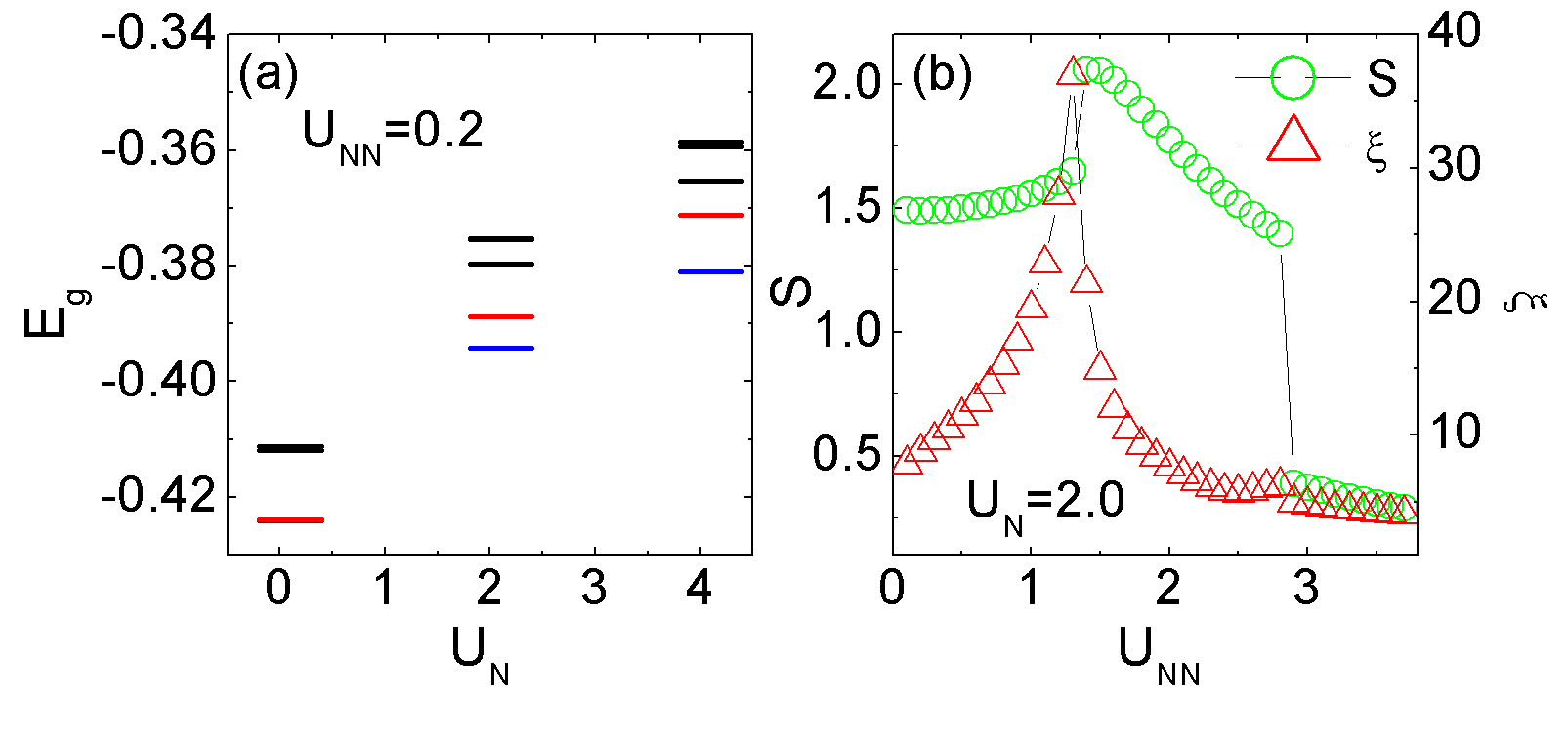}
 \caption{(a) displays the ED results of the lowest six eigenlevels of a half-filling 4$\times$3 lattice for $U_{\text{\tiny{N}}}=$0.1, 2.1, 4.1 with $U_{\text{\tiny{NN}}}=0.2$, of which the blue lines denote the ground energy level and the red lines indicate the first excitation energy. (b) plots the von-Neumann entropy and the correlation length of the system with respect to ${U_\text{\tiny{NN}}}/U_{\text{\tiny{N}}}$ with $U_{\text{\tiny{N}}}=2.0$. The data were obtained by iDMRG calculations on an infinite cylinder with MPS unit cell of size $4\times4$ and the bond dimension $\chi$ up to 200. \label{fig3}}
\end{figure}

Firstly, we determine the phase diagram of the TFB UHB model with respect to different $U_{\text{\tiny{N}}}$ and $U_{\text{\tiny{NN}}}$. Fig.~\ref{fig3} (a) displays the selected sets of energy spectrum of the TFB UHB model obtained by ED calculations. For conciseness, we only put the first lowest 6 eigenvalues for each pair of ($U_{\text{\tiny{NN}}}$, $U_{\text{\tiny{N}}}$) in the figure. We find that, when $U_{\text{\tiny{NN}}}\ll U_{\text{\tiny{N}}}$ and $U_{\text{\tiny{N}}}\leq1$, the ground energy is nearly double degenerated and this double manifold is separated by a moderate finite gap to the lowest exciting energy level. This kind of double degeneracy is a necessary condition for the occurrence of a FCI state and is consistent with the ED results of Ref[\cite{Wang2011}]. However, such a double degeneracy will be lifted by increasing either $U_{\text{\tiny{N}}}$ or $U_{\text{\tiny{NN}}}$, indicating the possible phase transitions.

Fig.~\ref{fig3} (b) presents the critical behaviors of the von-Neumann entanglement entropy and the correlation length for the TFB HUB model obtained by iDMRG calculations. With a fixed $U_{\text{\tiny{N}}}=2.0$, the system driven by the next-nearest neighbour coupling $U_{\text{\tiny{NN}}}$ exhibits two critical points: the first one at $U_{\text{\tiny{NN}}}=1.3$ is signified by a sharp peak of the correlation length and a clear jump of the von-Neumann entanglement entropy which implies a second-order quantum phase transition; by contrast, the correlation length near the other critical point at $U_{\text{\tiny{NN}}}=2.8$ shows only a jump behavior together with the von-Neumann entanglement entropy implying a first-order phase transition.

\begin{figure}
\includegraphics[width=1.0\linewidth]
{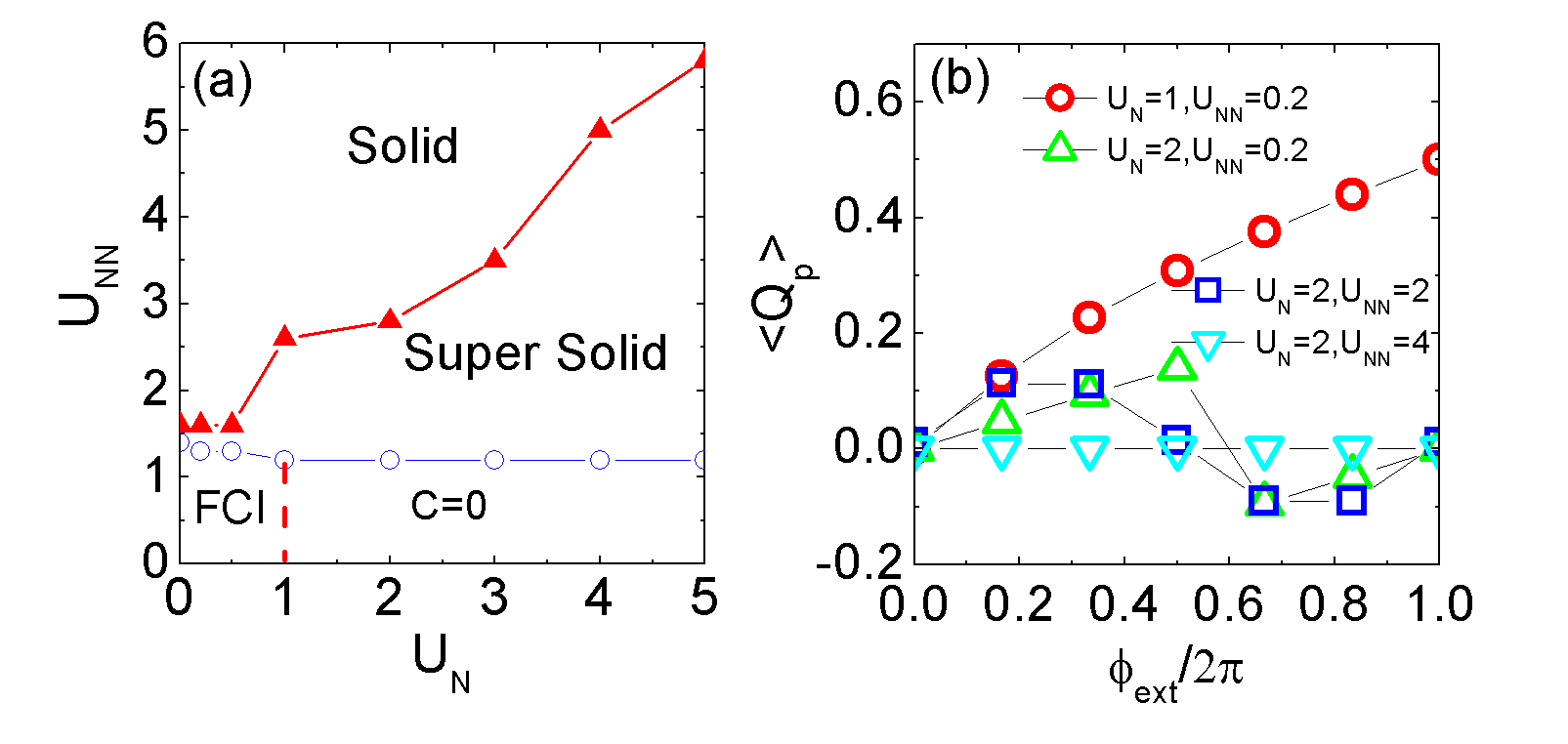}
 \caption{(a) the phase diagram of model~\ref{Hubbard1}. (b) the charge pumping with respect to the external flux $\phi_{ext}$ under distinct parameter settings. Here, we amplified the curve corresponding to the scae of $U_{\text{\tiny{N}}}=2$ and $U_{\text{\tiny{NN}}}=2$ by 1000 times in order to show its behavoir clearly.  All the data here were obtained by iDMRG calculations with an infinite cylinder with MPS bond dimension $\chi$ up to 600.  \label{fig4}}
\end{figure}

Fig.~\ref{fig4} (a) plots the phase diagram of the UHB model with respect to $U_{\text{\tiny{N}}}$ and $U_{\text{\tiny{NN}}}$ which is obtained by combining the data from Fig.~\ref{fig3} (a) and (b). There are four distinct phases: one FCI phase at the left bottom corner of the phase diagram, one non-degenerated topologically trivial phase labeled with $C=0$ beside the FCI phase, one super-solid phase and a solid phase. The topological nature of the FCI phase is of particular interest as shown below.

To determine the topological properties of the FCI region in Fig.~\ref{fig4} (a), we resort to the infinite density matrix renormalization group (iDMRG) method to study the charge pumping effect as of Laughlin's gedanken experiment\cite{Laughlin1981}. Technically, we add the external magnetic flux $\phi_{ext}$ threading through the cylinder which can be realized by imposing twist boundary conditions along the $y$ direction in the model during iDMRG calculation. Then, we adiabatically evolve the $\phi_{ext}$ from $0$ to $2\pi$ which means one has to calculate the ground-state wavefunction with respect to each value of the $\phi_{ext}$ in sequence by utilizing the wave function obtained in the last DMRG step as the initial trial wavefunction. After the adiabatic evolution and due to the particle hopping between the edge modes\cite{Alexandradinata2011} in the topological FCI phase, the ground state is restored to the one before the insertion of $\phi_{ext}$ but with a different particle distribution which leads to the passing of $<Q_p>$  particles across the middle cut point of the matrix product state (MPS) wave function on the cylinder; here $<Q_p>$ featuring the quantized Hall conductivity just give the Chern number of the topological phase and we find that it is exactly $1/2$ as illustrated in Fig.~\ref{fig4} (b) for the case of $U_{\text{\tiny{N}}}=1$, $U_{\text{\tiny{NN}}}=0.2$. This demonstrates the existence of a FCI phase in the TFB UHB model.

\begin{figure}
\includegraphics[width=1.0\linewidth]
{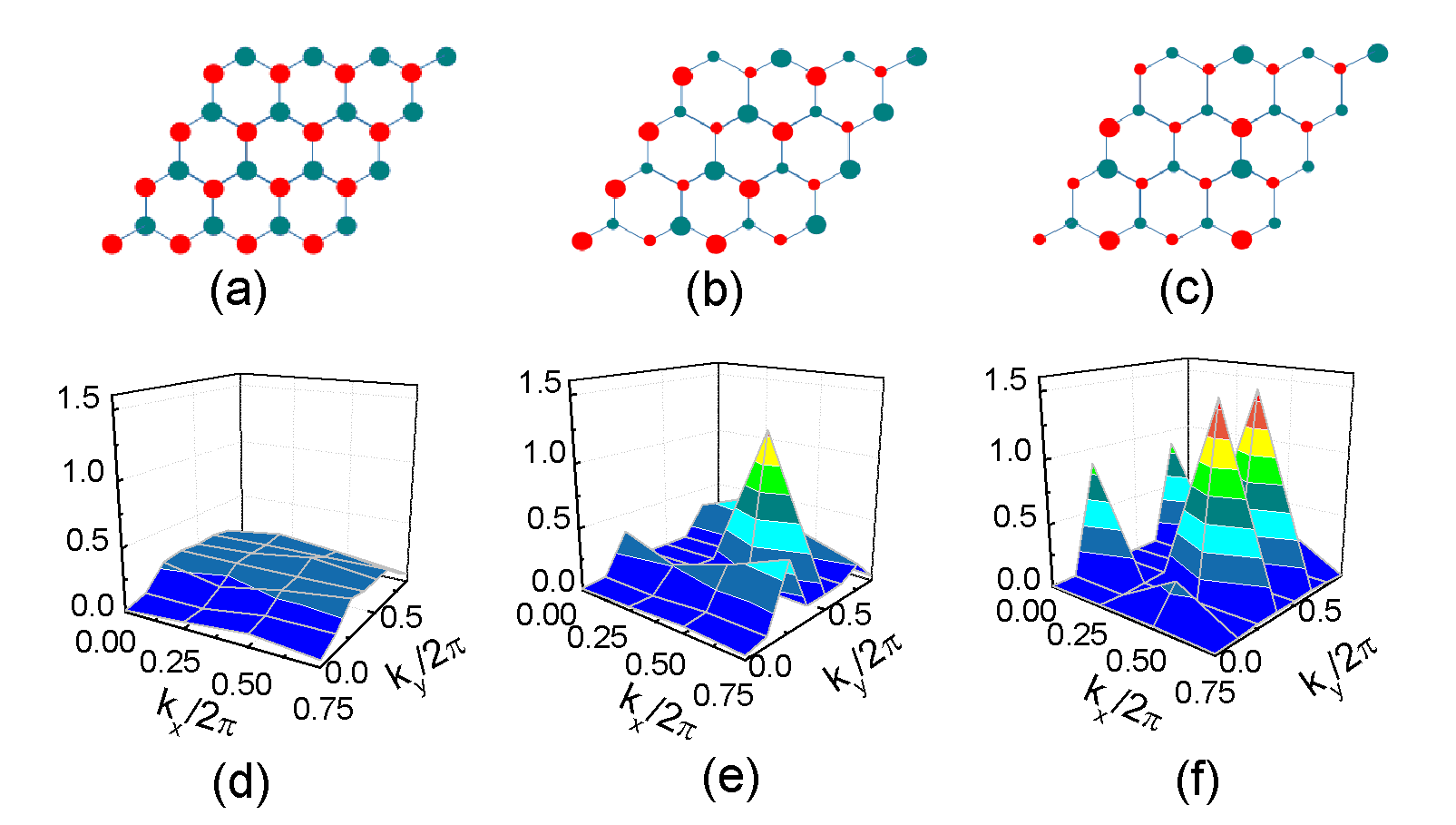}
 \caption{The corresponding on-site charge distribution in the phase diagram. Here, red and blue dots denote different sub-lattices and the magnitude of the on-site charge density is signified by the diameter of the dots. (a): the FCI phase and the topological trivial phase with $C=0$ with $U_{\text{\tiny{N}}}=2$ and $U_{\text{\tiny{NN}}}=0.2$ (b): Super-solid phase with $U_{\text{\tiny{N}}}=2$ and $U_{\text{\tiny{NN}}}=2$ (c): Solid phase with $U_{\text{\tiny{N}}}=2$ and $U_{\text{\tiny{NN}}}=4$. The data were obtained from iDMRG calculations. \label{fig5}}
\end{figure}

The stability of the FCI state with respec to internal or external factors has been investigated in many previous works, e.g. Ref.\onlinecite{Andrews2021}; in the article, we also ascertain that the FCI phase is not stable in the presence of the nearest-neighbouring coupling $U_{\text{\tiny{N}}}=1$. As shown in Fig.~\ref{fig4} (a), the FCI state is confined in the weak-coupling regime in the phase diagram. Fig.~\ref{fig3} (a) reveals that the double ground state degeneracy can be broken up by a larger $U_{\text{\tiny{N}}}>1$ which is further supported by the absence of charge pumping effect as shown in Fig.~\ref{fig4} (b) for the case of $U_{\text{\tiny{N}}}=2$, $U_{\text{\tiny{NN}}}=0.2$. However, this topological-non-topological transition is difficult to be detected by using the criteria of correlation length or von-Neumann entanglement entropy since they do not exhibit any singularities near the critical point; the ground-state on-site particle occupancy pattern for these two phases shows uniform charge distribution with an on-site charge density of $\langle n_i\rangle=1/4$ as shown in Fig.~\ref{fig5} (a). Hence, we can not exactly fix the phase boundary between these two phases but roughly mark it by the absence of Chern number $C=0$ in Fig.~\ref{fig4} (a) according to the occurrence of the charge pumping phenomena in Fig.~\ref{fig4} (b) and the double ground state degeneracy in Fig.~\ref{fig3} (a).

By increasing $U_{NN}$, we also find two topological trivial phases: the super-solid phase and the solid phase as illustrated in Fig.~\ref{fig4} (a) which is very analogous to the scenario in \onlinecite{Luo2020}. The on-site particle density and the corresponding structure factor $S(\mathbf{q})$ for these two phases are shown in Fig.~\ref{fig5} (b) and (c). The super-solid phase is characterized by the $(\pi,\pi)$ peak of $S(\mathbf{q})$ and the sinusoidal pattern of charge pumping as shown in Fig.~\ref{fig4} (b). The super-solid phase is driven into the solid phase by increasing $U_{NN}$ across a first-order phase transition as indicated in Fig.~\ref{fig3} (b). In such a solid phase, the charge pumping effect disappears as shown in Fig.~\ref{fig4} (b) and a large gap shows up in the entanglement spectrum above the lowest energy of which the data is not shown here. A difference to \onlinecite{Luo2020} is that the two peaks of $S(\mathbf{q})$ at $(\pi,\pi/2)$ and $(\pi,3\pi/2)$ are shifted to $(\pi,\pi/4)$ and $(\pi,3\pi/4)$ which is due to the physical effect of the special complex hopping phase factors of the UHB model. The conclusions about these two phases are almost the same as the results of \onlinecite{Luo2020} and no more discussions are necessary to be made in this article.
\section{Conclusions}\label{sec04}
In summary, we propose a method to simulate a fractional Chern insulator based on Rydberg-dressed neutral atoms. By using Rydberg dressing, a hard-core boson Hubbard model on a honeycomb lattice can be realized and in which the single-particle ground energy band can be described by a factional Chern number of $1/2$ concerning their topology. The topological states persist in the presence of weak many-body interactions , which we have demonstrated by ED and iDMRG calculations. The results show that anomalous fractional quantum Hall phases can be emulated with neutral atoms via Rydberg dressing.

However, it should be noted that it is not easy and direct for us to determine the parameter settings for experimental groups to prepare the topological flat band of the UHB model. Our results in this paper only give a prediction of the possible ranges for the parameters that may be used to realize the TFB UHB model. Further effort will be devoted to designing a more promising experimental precedure to construct such a model; the relating results will be shown in our future studies.
\section*{acknowledgements}\label{sec05}
We thank Shou-Shu Gong and Wei-Wei Luo for very useful discussions. Yang Zhao acknowledges support from the Natural Science Basic Research Plan in Shaanxi Province of China under Grant No. 2021JM-041, and X.-F. Shi is supported by the National Natural Science Foundation of China under Grants No. 12074300 and No. 11805146, and the Fundamental Research Funds for the Central Universities. The iDMRG calculations were performed using the TeNPy Library\cite{tenpy}.

\bibliography{TopologicalInsulator}
\end{document}